\documentclass{svproc}
\usepackage{url}

\usepackage{xcolor}
\usepackage{framed}
\usepackage[normalem]{ulem}
\usepackage{enumerate}
\usepackage{algorithmic}
\usepackage[ruled, vlined,noend]{algorithm2e}
\usepackage{wrapfig} 
\usepackage{caption}
\usepackage{subcaption}
\usepackage{xcolor}
\usepackage{hyperref}

\usepackage{amsmath}
\usepackage{bbm}

\usepackage{tikz}
\usepackage{verbatim}
\usetikzlibrary{arrows,calc}
\usepackage{relsize}

\usepackage{amsmath}

\begin{document}
\mainmatter              % start of a contribution
\title{A Matching Mechanism for Provision of Housing to the Marginalized}
\titlerunning{A Matching Mechanism for Provision of Housing to the Marginalized}  % abbreviated title (for running head)
\author{J Ceasar Aguma}
\authorrunning{J Ceasar Aguma} % abbreviated author list (for running head)
\institute{University of California Irvine, Irvine CA, USA,\\
\email{jaguma@uci.edu},\\ WWW home page:
\texttt{https://jaguma.wixsite.com/j-ceasar-aguma}}

\maketitle              % typeset the title of the contribution

% \begin{abstract}
% The abstract should summarize the contents of the paper
% using at least 70 and at most 150 words. It will be set in 9-point
% font size and be inset 1.0 cm from the right and left margins.
% There will be two blank lines before and after the Abstract. \dots
% % We would like to encourage you to list your keywords within
% % the abstract section using the \keywords{...} command.
% \keywords{computational geometry, graph theory, Hamilton cycles}
% \end{abstract}
\begin{abstract}
During this pandemic, there have been unprecedented community and local government efforts to slow down the spread of the coronavirus, and also to protect our local economies. One such effort is California’s project Roomkey that provided emergency housing to over 2,000 vulnerable persons but fell short of the set goal of 15,000. It is projected that the homelessness problem will only get worse after the pandemic. With that in mind, we borrow from efforts like project Roomkey and suggest a solution that looks to improve upon these efforts to efficiently assign housing to the unhoused in our communities. The pandemic, together with the project Roomkey, shed light on the underlying supply demand mismatch that presents an opportunity for a matching mechanism solution to assigning housing options to the unhoused in a way that maximizes social welfare and minimizes susceptibility to strategic manipulation. Additionally, we argue that this automated solution would cut down on the amount of funding and personnel required for the assignment of housing to unhoused persons. Our solution is not intended to replace current solutions to homeless housing assignments but rather improve upon them. We can not postpone a proper solution to homelessness anymore, the time is now as the need for an efficient solution is most dire.  

\end{abstract}
\section{Introduction}

In this global pandemic, humanity as a collective has been awakened to what is most important to our unified survival. Now more than ever, we understand the significance of a permanent shelter to call home. However, while many of us could stay indoors, and protect ourselves and our communities from the spread of the virus, those unhoused among us were, and still have been left vulnerable. The United States Department of Housing and Urban Development reported the homeless population to be over 500,000 across the nation\cite{hudData}. Of these 500,000, Culhane et al. estimate the modal age to be between 50-55 in several cities\cite{culhane}. This happens to be the most COVID-19 vulnerable group as reported by the Center for Disease Control(CDC). In what further emphasize this vulnerability, a 2019 study found that 84\% of the unhoused population self-reported to have pre-existing physical health conditions\cite{rountree}. California and New York, states that have been gravely affected by COVID-19, also have the largest unhoused populations. This summary does not even tell the global story which paints an even bleaker picture. 

Los Angeles and many other cities scrambled to provide temporary housing for the unhoused during the pandemic through tent cities and vacant hotel rooms\cite{holland}. However, most of these were either poorly assigned, as in the case of disabled persons, or left vacant because of the lack of an efficient allocation procedure. This metropolitan effort also leaves a few questions unanswered, for example, what happens after this pandemic? How many other people will be left unhoused? How many additional housing options will become available for low-income persons? To answer some of these questions and meet the need for a better housing assignment procedure, we propose a matching mechanism to improve the allocation of available housing to unhoused marginalized groups such as veterans and low-income families. In the background, we review LA county's project Roomkey initiative and set the stage for a matching mechanism that could improve this initiative.

\subsection{Background}

\subsubsection{Case Study: Project Roomkey}

According to the LA county COVID-19 website, project Roomkey is "a collaborative effort by the State, County, and the Los Angeles Homeless Services Authority (LAHSA) to secure hotel and motel rooms for vulnerable people experiencing homelessness. It provides a way for people who don’t have a home to stay inside to prevent the spread of COVID-19" \cite{laCounty}. Eligible persons, where eligibility is determined on the basis of vulnerability to COVID-19 and a reference from a local homeless shelter or law enforcement office, are assigned temporary housing in the form of hotel and motel rooms.

\par The matching of eligible persons has been done by local homeless shelters that match the individuals to available hotel or motel rooms in their locality. Whether this is automated is unclear, but given the program's failures, one would assume that the matching was NOT done by a central clearing house but rather arbitrarily without full knowledge of preferences and optimal matches.
\par Furthermore, the program was not clear on how individuals would be moved to permanent or transitional housing when it closes. To quote the website, "while participants are staying at these hotels, on-site service providers are working with each client individually to develop an exit plan, with the goal of moving them to a situation that permanently resolves their homelessness. In cases where this isn’t feasible, LAHSA will use existing shelter capacity to move people into an interim housing environment or explore other options"\cite{laCounty}. The key part is "on-site service providers are working with each client individually," which implies that the matches are not automated and were made depending on whatever information was locally-available to the on-site service provider. The LA Times has highlighted some failures in the project, for example, the project was slammed for discriminating against the elderly and disabled because," the agency deliberately excluded those who cannot handle their own basic activities, such as going to the toilet or getting out of bed"\cite{laroomkey}. The project's leadership cited a lack of personnel and funding as the reason it did not succeed\cite{laroomkey1}. So clearly, a cheap and automated option for matching individuals to housing options is required.

\par The project is now coming to an end after housing about 30\% of the projected total. While the program has been reported as a failure, it allows one to imagine a real solution to homelessness in LA county and in fact, any metropolis. What the program showed is that there is room for a central matching mechanism that can help move persons from the streets into shelters and from shelters into permanent housing. What we will show below, is that this mechanism can be designed to be Pareto-optimal (assign every person their best possible option at the time of assignment), and strategy-proof (persons cannot do better by cheating in this mechanism). Given a lack of funding and personnel, we felt that an automated matching mechanism that is theoretically optimal would be a great solution.

\par Further analysis of project Roomkey reveals a rich structure that reinforces the need for a matching mechanism. We will give a detailed look at this structure in a later section, and only a summary here. Because of state and federal mandated lockdowns, hotels and motels found themselves with large volumes of vacant rooms, an oversupply of sorts. In the same communities as the oversupplied hotel and motel rooms are the many unhoused folks that, due to different circumstances, can not afford to access and pay for the vacant rooms but do demand shelter, more so in a pandemic with federal and state-mandated lockdowns. What we see here is an oversupply of a commodity/service, and an abundance of demand but the two sides are inaccessible to each other without the help of a third party like the local, state, or federal government. This third party is what we consider as the matching mechanism designer, something, we will argue, should have done better when matching unhoused folks to the vacant rooms. This text, therefore, intervenes at this point, to further highlight the structure of oversupply to handicapped demand, and calling for a simple but sophisticated matching mechanism that can navigate locality constraints that arose in the allocation of vacant rooms to unhoused persons.

%\subsubsection{Case Study: VASH}
%%HUD-VASH is a collaborative program between HUD and VA that combines HUD housing vouchers with VA supportive services to help Veterans who are homeless and their families find and sustain permanent housing. Through public housing authorities, HUD provides rental assistance vouchers for privately owned housing to Veterans who are eligible for VA health care services and are experiencing homelessness. VA case managers may connect these Veterans with support services such as health care, mental health treatment, and substance use counseling to help them in their recovery process and with their ability to maintain housing in the community. Among VA homeless continuum of care programs, HUD-VASH enrolls the largest number and largest percentage of Veterans who have experienced long-term or repeated homelessness. At the end of FY 2019, there were 90,749 Veterans with active HUD-VASH vouchers and 83,684 vouchers in use.%%

%%VA eligibility program % 
%Over 14,000 vouchers went unused in 2019, with the no. of homeless vets still roughly 38,000. The VA also cited a lack of personnel and rising rent costs across the country (another way to say lack of funding) as the reasons why these vouchers went unused.(militarytimes paper)%

\subsubsection{Literature Review}
This paper contributes to a well-established body of work on homelessness, matching markets applied towards social good and matching mechanisms specifically for housing assignment. Below we review a few key papers on the above-mentioned research topics.
\par While we highlight the need for a matching mechanism to mitigate homelessness in cities around the globe, there is a long history of scholars deploying matching or mechanism design towards efficient housing solutions. We will summarize some relevant and notable works here. 
\par Theoretical scholars have been studying housing matching markets from as far back as 1974 when Shapley and Scarf put forth economic mechanism theory for the housing market with existing tenants and introduced the Gale Top Trading Cycles algorithm\cite{SHAPLEY197423}. In 1979, Hylland and Zeckhauser set the foundation for a house allocation problem with new applicants, defining a housing market core\cite{hylland}. Abdulkadiroglu and Sonmez extend the work to a model with new and existing tenants\cite{sonmez}. We direct the reader to \cite{abdulkadiroglu2013matching} for a more comprehensive review of matching markets theory.
\par O’Flaherty goes beyond economic game theory to provide a full economic “theory of the housing market that includes homelessness and relates it to measurable phenomena"\cite{OFLAHERTY199513}. He later extends the work to answering “when and how operators of shelters should place homeless families in subsidized housing”\cite{OFLAHERTY200969} and also updates the economics of "homelessness under a dynamic stochastic framework in continuous time”\cite{OFLAHERTY201277}.
\par Sharam gives a comprehensive breakdown of how matching markets have been applied towards the provision of new subsidized multifamily housing for low-income families in Australia\cite{Sharam}. Sharam also illustrates ways in which the use of digital platforms for matching could help improve the optimality of matching in housing assistance\cite{Sharam2}. To the best of our knowledge, \cite{Sharam} and \cite{Sharam2} are the only texts that explore the use of a matching mechanism for the provision of low-income housing. Sharam, however, does not extend the work to marginalized groups and only considers the Australian effort whereas we look to create a mechanism that not only looks at low-income multifamilies but all unhoused persons.
\par Because of the overwhelmingly unstable labor and the housing market at the present, Hanratty’s work on the impact of local economic conditions on homelessness using Housing and Urban Development(HUD) data from 2007-2014\cite{hanratty} is also very relevant to our research. Mansur et al., which examines policies to reduce homelessness, will be useful for the future work in this research that concerns itself with policy recommendation\cite{MANSUR2002316}. 

\subsubsection{Our contribution}
Building from all this past work and the unique structure of the project Roomkey, we provide a matching mechanism for better interim and/or permanent housing assignments for unhoused individuals. This mechanism is derived from those explored in \cite{SHAPLEY197423},\cite{hylland}, and \cite{sonmez}. We show that this mechanism is also Pareto-optimal and strategy-proof. We also present a clear picture of that unique structure underlying project Roomkey and invite scholars to pay attention to other areas where this structure presents itself, for example, the food industry.

\section{Model}

% \begin{table}[!htbp]\caption{Table of Notation}
% \begin{center}% used the environment to augment the vertical space
% % between the caption and the table
% \begin{tabular}{r c p{6cm} }
% \toprule
% $n$ & $\triangleq$ & number of persons $i$ \\
% $m$ & $\triangleq$ & number of housing options $j$ \\
% $(\pi_{i}(j))$ & $\triangleq$ & person $i$'s preference list \\
% $R$ & $\triangleq$ & priority ranking $R$ \\
% \bottomrule
% \end{tabular}
% \end{center}
% \label{tab:TableOfNotationForMyResearch}
% \end{table}

\subsection{Problem Formulation}

\par Consider a metropolis where $n$ agents, which would be persons without permanent housing, ranging from multifamilies to single individuals entered in a shelter or veteran affairs database, looking to transition to better housing options. Let us further assume a collection of available housing options $m$ in many forms: low rent apartments, vacant motels or hotels, tent cities, or group homes (many housing options of this kind have been acquired or created by local governments during the COVID-19 lockdown). A person $i$ has a preference list $(\pi_{i}(j))$ on the housing options which is derived from their individual preferences on size, location, cost, accessibility, and many others. We assume that there is no preference list on the persons as that could open the model up to circumstantial bias. However, we assume that there exists a priority ranking, $R$ on the agents on basis of factors like family size, health risk like the current COVID-19 risk for elders, time spent waiting for a housing assignment. How $R$ is determined is left up to the decision maker like the Veteran Affairs office or city officials. (In fact, most shelter or housing authorities almost always have such priority lists already. For example for COVID-19 emergency housing, persons under the highest COVID-19 risk were given top priority.)
%  \begin{figure}[!htbp]
%     \begin{center}
%         \includegraphics[width=1.0\textwidth]{images/model.png}
%         \caption{The housing matching model}
%         \label{fig:model}
%     \end{center}
% \end{figure}
\subsubsection{Research goal}
\par The goal then is to design a mechanism that matches the $n$ persons to the $m$ housing options according the preference lists and priority ranking while considering that $n = m$, $n > m$, or $n < m$. Because we intend to implement this model with local policy makers, we have the additional goals of minimizing cost of implementation and personnel required.

% \subsubsection{Generating a preference list}

With a model defined and a research goal specified, the next section details how this goal is attained using a simple matching algorithm.

\subsection{The Matching Mechanism}
\par We have established that there three cases expected in this matching problem. We will give an algorithm for each one of these here, starting with the most straightforward case where $m = n$, then explain how simple modification can help tackle the other two cases.

\begin{algorithm}
	\SetAlgoNoLine
	Organize agents in some priority queue in descending order (ties are broken randomly)\newline
    \For{Each agent in the queue}{
    \begin{enumerate}
        \item Assign them the best housing option currently available according to their preference list
        \item Terminate when queue is empty 
    \end{enumerate}}
\caption{A matching algorithm for assigning housing to the marginalized when $m = n$}
	\label{ALG}
\end{algorithm}

\par Quick inspection will reveal from in the literature of matching markets, this algorithm is in fact serial dictatorship with a fixed priority queue in place of a random order on agents. Like random serial dictatorship, this algorithm is Pareto-optimal and strategy proof. For completeness, proofs for both will be given here.
\section{Analysis}

\par For the case of $n > m$, the last $n - m$ agents in the priority queue simply maintain their current housing options. So, in a sense, we tackle this case the same way we go about the $m = n$ case.
This is also true for $m > n$, where the $m - n$ least preferred housing options are simply left unassigned. A definition and proof for Pareto optimality follows.

\subsection{Pareto optimality}
As a precursor to the proof, we provide a definition of Pareto Optimality.
\par Given $n$ users and $n$ resources, an assignment \newline
$X = (x_{1},x_{2},........,x_{n-1},x_{n})$ is Pareto optimal, 
if it is not Pareto dominated by any other assignment \newline
$X' = (x'_{1},x'_{2},........,x'_{n-1},x'_{n})$.

Assignment $X'$ Pareto dominates $X$ if for each user $i$; $$x_{i} \succeq x'_{i}$$ with at least one user $j$ for whom ; $$x_{j} \succ x'_{j}$$

\begin{theorem}
The simple Algorithm 1 above is Pareto optimal in all three cases; $m<n, m=n,$ and $ m>n$.
\end{theorem}

\subsubsection{Proof}
Let us assume that $X$ is not Pareto-optimal. This means $X$ is dominated by another matching assignment $X'$ in which at least one agent $j$ must have a better and different allocation $a$. But we know that $X$ assigns every agent their best available preference at the time of assignment. So if $j$ indeed has a better assignment in $X'$, this would mean that an agent $i$ (who got the assignment $a$ in $X$) earlier in the priority queue also has a different assignment in $X'$. Observe that agent $i$'s assignment is either worse in $X'$ or must be an assignment that was awarded to another agent earlier in the priority queue in $X$. One can follow this cycle until at least one agent gets a worse off assignment in $X'$. This presents a contradiction because now we see that either $j$ does better and another agent does worse in $X'$ or $j$ themselves gets a different option that is not their best available option, in which case they would do worse in $X'$. Therefore, it is impossible that $X'$ dominates $X$. To illustrate this better, we provide a few examples below.

\subsubsection{Example 1: $m = n$}
Given an agent set ${i,j,k}$, and housing options ${a,b,c}$. With a priority queue: $i-j-k$ and  preference lists: 
\begin{gather}
i: a \succ b \succ c \\ j: b \succ c \succ a \\ k: c \succ a \succ b 
\end{gather}
Our algorithm would assign housing options as follows, ${x_i = a, x_j = b, x_k = c}$. All three agents would get their best options and so any other algorithm must either produce the same assignment or at least one agent would be worse off.
\par if we altered the preference lists to: 
\begin{gather} 
i: a \succ b \succ c \\ j: a \succ c \succ b \\ k: c \succ a \succ b 
\end{gather}

Our algorithm would assign housing options as follows, ${x_i = a, x_j = c, x_k = b}$. Observe that in this case $j$ and $k$ do not get their best possible assignment but get the best available assignments. If another algorithm gave $j$ housing option $a$, then $i$ must get a different assignment and hence be worse off.

\par We ask the reader to try out different permutations of the preference lists and check to see that in each, no other assignment would dominate that produced by algorithm 1 given in this text.

\subsubsection{Example 1: $m < n$}
\par For the case where, $m < n$ , the same algorithm is employed but this time the $n- m$ remaining agents in the priority queue simply maintain their existing housing options or, more harshly put, do not get a new housing option.
\par Consider an agent set ${i,j,k,l}$, and housing options ${a,b,c}$. With a priority queue: $i-j-k-l$ and  preference lists: 
\begin{gather}
i: a \succ b \succ c \\ j: b \succ c \succ a \\ k: c \succ a \succ b \\ l: a \succ c \succ b 
\end{gather}
Our algorithm would assign housing options as follows, \newline ${x_i = a, x_j = b, x_k = c, x_l = None}$. Any algorithm that gives $l$ any of the options $a,b,c$ would leave another agent worse off, unless the number of housing options increased.

% \subsubsection{Example 1: $m > n$}
% \par For the case where, $m > n$ , we can quickly reduce this down to $m = n$ by eliminating the least preferred $m - n$ housing options.
% \par Consider an agent set ${i,j,k}$, and housing options ${a,b,c,d}$. With a priority queue: $i-j-k-l$ and  preference lists: 
% \begin{gather}
% i: a \succ b \succ c \succ d \\ j: d \succ c \succ a \succ b \\ k: c \succ a \succ d \succ b 
% \end{gather}
% Our algorithm would assign housing options as follows, ${x_i = a, x_j = d, x_k = c}$ eliminating option $b$. This is still a Pareto-optimal assignment. Again, we ask the reader to cross check other preference list permutations.

\subsection{StrategyProofness}
\par A matching mechanism is strategy-proof if truth telling is a utility-maximizing strategy, that is, the only way an agent can be guaranteed to get their best possible assignment is if they report true information.

\begin{theorem}
Algorithm 1 is strategy-proof because the only way an agent gets their best option is by picking it in their turn.
\end{theorem}

\subsubsection{Proof}
If we assume that the priority queue is out of the agents' control and decided by a third party like a policy-making entity or shelter management according to some standard criteria like period of time spent waiting for a housing option, then it's easy to see that agents can not cheat by misreporting their preferences. The only way an agents their best option is if this option is correctly placed in their preference order and is available when their assignment turn comes.
\par \textbf{A quick example}; let us assume we have three agents ${i,j,k}$ with that exact order in the priority queue,i.e; $i-j-k$. With housing options ${a,b,c}$, lets also assume their true preferences are as follows; 
\begin{gather}
i: c \succ a \succ b \\ j: c \succ b \succ a \\ k: b \succ a \succ c 
\end{gather}
The algorithm would assign housing options as follows, ${x_i = c, x_j = b, x_k = a}$ (The reader can check for Pareto optimality).
But if agent $j$ misreports their preferences as $j: a \succ b \succ c$, they would get $a$ when their best option $c$ was available. In fact, if $j$ alters their preference list in any way, they would get most likely miss out on getting their best option. We, therefore, can say this algorithm 1 is strategy-proof.

\section{Project RoomKey Revisited}
% Summarize LAHSA alg and show Pareto deficiency and strategy unproofness.

In this section, we will investigate how the algorithm proposed by this text compares to the current algorithm employed by LAHSA for assigning housing under project \newline Roomkey. Of course, LAHSA does not officially call their procedure an algorithm or have a clear outline of steps taken in assigning housing options. We had to read through their program policies and procedures \cite{lahsa} and decipher some outline of the implicit algorithm they use for the assignments. 
\par Below, we will present preliminaries to that algorithm including the non-trivial structure that allowed for the proposal of project Roomkey, the project's priority criteria, the algorithm itself, and an analysis of it.

\subsection{The Over Supply, Low-income demand picture}
As a precursor to further evaluation of project Roomkey, we would like to highlight the unique structure that rendered project Roomkey necessary. However, this is simply an introduction of this structure, more elaborate treatment of the structure will be done in future work. This structure is nevertheless crucial to understanding the matching problem that made project Roomkey necessary.
 \begin{figure}[!htbp]
    \begin{center}
        \includegraphics[width=1.0\textwidth]{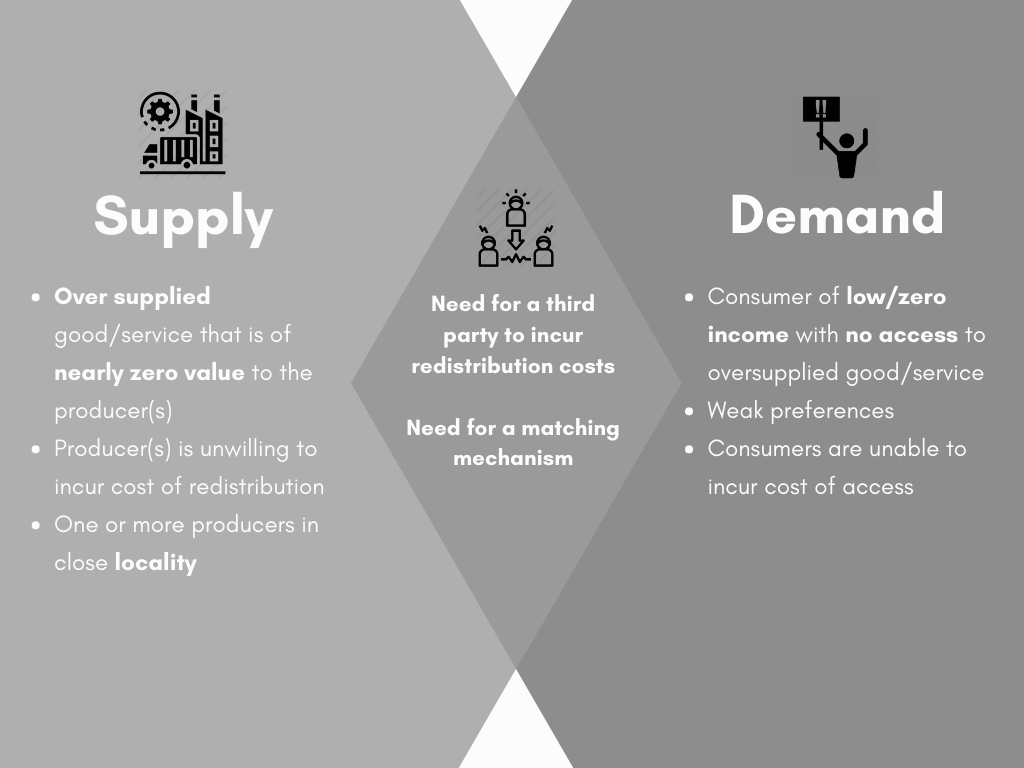}
        \caption{The unique 'unmatched' structure}
        \label{fig:unmatched}
    \end{center}
\end{figure}

Figure 2 illustrates the two sides that create the matching scenario. We have a producer that has an \textbf{oversupply of a commodity or service} , and because of the oversupply, the commodity/service is of \textbf{nearly zero value} to them. The pandemic created this situation for hotel and motel owners who suddenly had an abundance of rooms, that in many places around the world, were left unused. 
\par Adjacent to this is the demand that by different circumstances, is rendered unable to access the oversupplied commodity/service. Circumstances like low to zero income to purchase a commodity that the producer would rather waste than avail for cheap or charity. In many cities around the world, unhoused persons demand these rooms but can not access them because there is rarely a third party (or producer) willing to incur the cost of redistribution. This is the scenario that created the vacuum for project Roomkey to fill.
\par With California state and local governments stepping in to incur the cost of redistribution, this matching of oversupplied commodities/services to handicapped demand happens. The one step left to reconcile then is, how to efficiently match the vacant rooms to the unhoused folks. Below, we will compare project Roomkey's matching procedure to the one proposed by this text, in the context of Pareto optimality and strategyProofness.

\subsection{Project Roomkey Housing assignment}
From \cite{lahsa}, given a $n$ unhoused persons and $m$ housing options \textbf{distributed among different homeless service provides}, the algorithm for assignment is as follows:
\begin{algorithm}
	\SetAlgoNoLine
	Organize agents in some priority list \newline
    \For{Each eligible agent on the list}{
    \begin{itemize}
        \item Assign agent to a local homeless service \newline provider
        \item The local homeless service provider assigns \newline the agent a housing option according \newline to their needs
    \end{itemize}}
\caption{The assignment procedure employed by LAHSA under project Roomkey}
	\label{ALG2}
\end{algorithm}

Eligibility and priority for assignment of interim housing under \newline project Roomkey are determined by "high-risk profile for COVID-19" \cite{lahsa}. According to LAHSA, high-risk is defined or determined by \textbf{age, chronic health condition, COVID-19 asymptomatic condition, persons currently staying in congregate facilities}. A priority list is generated from the above criteria\cite{lahsa}.

The immediate red flag from this algorithm is that preferences and assignment are restricted by locality from the fact that the $m$ housing options are distributed among local homeless service providers. Better options according to one's preferences could exist through another local service provider but they would never be available to this individual. We will do a deeper analysis of the above algorithm (check for Pareto optimality and strategy proofness) next.

\subsection{Pareto Optimality}
In the proof for why algorithm 1 is Pareto optimal, we showed two examples where the algorithm always finds an assignment that can not be improved without making any agent worse off, we then dared the reader to find an example that proves otherwise. Here we will show an example in which algorithm 1 dominates LAHSA's algorithm 2. This is sufficient to prove that algorithm 2 is not Pareto optimal.

\subsubsection{Example: Two homeless service providers}
We will assume that the LAHSA has two local homeless service providers in different localities under the project Roomkey. Homeless service provider, $P$ has housing options, ${a,b,c}$ available. While homeless service provider, $Q$ has housing options, ${x,z}$. We additionally assume an eligible person $i$ seeking a housing option with the following preference list generated from their needs: $$i: z \succ b \succ c \succ x \\ $$

Under algorithm 2, we have two possible outcomes that depend on whether the LAHSA sends $i$ to $P$ or $Q$. 

\begin{itemize}
    \item If $P$, then $i$ will most likely be assigned housing option $b$
    \item And if $Q$, then $i$ probably gets their most preferred option $z$
\end{itemize}

Observe that because of locality, $i$ could be assigned to an option that does not best fit their needs. This means that there exists cases where algorithm 2 can be dominated by another algorithm that can guarantee a better housing option to $i$. 
\par One such algorithm is algorithm 1 where we would have all the available housing options ${a,b,c,x,z}$ in one database. We would then assign $i$ their most preferred housing option $z$.
Algorithm 1 clearly Pareto dominates algorithm 2 in this example. As a counter-example, one could ask, what if there was another person $j$ in the locality of $Q$ that also preferred $Z$? Assigning $z$ to $i$ would sure leave $j$ worse off. Let us set up this example and see why this is not a valid counter-example.

\subsubsection{Example: Two homeless service providers and two eligible persons}
We will assume that the LAHSA has two homeless service providers in different localities under the project Roomkey. Homeless service provider, $P$ has housing options, ${a,b,c}$ available. While homeless service provider, $Q$ has housing options, ${x,z}$. We additionally assume two eligible persons $i \& j$ seeking housing with the following preference list generated from their needs:
\begin{gather} 
i: z \succ b \succ c \succ x \\ j: z \succ a \succ c \succ b 
\end{gather}

Under algorithm 2, we assume that LAHSA sends both to the homeless service provider of their respective locality, that is, $i$  to $P$  and $j$  to $Q$, $j$ would then be assigned their most preferred option $z$ but $i$ would get $b$. However, if $i$ has higher priority, then we see that algorithm 2 would never properly honor that priority while algorithm 1 would rightfully assign $z$ to $i$  and $a$ to $j$ which are their best possible outcome given a descending priority ordering of $i-j$.

\subsection{Strategy Proofness}
Example 4.3.2 demonstrates that it would be possible for someone to get a better housing option by simply misreporting their locality to LAHSA. We present that example again here, with a few changes, as a proof that algorithm 2 is not strategy proof.

\subsubsection{Example: Two homeless service providers and two eligible persons revisited}
We will assume that the LAHSA has two homeless service providers in different localities under the project Roomkey. Homeless service provider, $P$ has housing options, ${a,b,c}$ available. While homeless service provider, $Q$ has housing options, ${x,z}$. We additionally assume two eligible persons $i \& j$ in the localities of $P \& Q$ respectively, seeking housing with the following preference list generated from their needs: 
\begin{gather} 
i: a \succ z \succ c \succ x \\ j: a \succ x \succ c \succ b 
\end{gather}
Under algorithm 2, If $j$ misreported their locality, they would have a shot at getting their most preferred option $a$, but if the priority queue of $i - j$ is followed, they would end up with $x$.
\par Observe that under algorithm 1, it would not matter if $j$ misreports or not, they would get option $x$ either way while $i$ will always get option $a$ (which would be their rightful assignments according to the priority queue).
\par This simple example shows us that, indeed, persons can improve their chances by misreporting their preferences and locality in algorithm 2, which could leave them worse off whereas algorithm 1 protects against such incidents.

\subsection{Locality Expansion}

\par Locality is a key component in the comparison of our matching scheme to that employed by project Roomkey. And by locality, we mean the area considered when assigning housing option to new persons. If $U(p)$ is the expected utility for an individual $k$ for an allocation from the $m$ housing options, with a utility $u(x_i)$ and probability $p_i$ for each housing option, we get the following definition.
$$
U(p) = \sum_{i=1}^{m} u(x_i)p_i
$$
It is easy to see that the expected utility for any individual $i$ is nondecreasing with increase in  size of the locality or the number of housing providers as long as all the utilities are nonnegative.

\section{Conclusion}
\subsection{Discussion}
For several decades now, matching mechanisms have been deployed, sometimes invincibly, to solve economic and social problems with unprecedented efficiency. Most notably, the kidney transplant matching algorithm was key to saving 39,000 lives in 2019 alone\cite{kidneyMatch}. Like past matching mechanism solutions to social problems, the proposed mechanism promises to solve an age-old, complex problem more efficiently. With a Pareto optimal algorithm, we have confidence that this solution would be fair and consequentially improve social welfare without being susceptible to unfair strategies from those trying to cheat their way in.

\par Automating the assignment of housing options, which is currently done one-on-one by local service providers, should help reduce the amount of personnel required by undertakings like LA county's project Roomkey that cited a lack of sufficient personnel as a hindrance to its success. Besides personnel, we speculate that automation would also render other parts of the current system obsolete therefore resulting in a reduction of cost. This, too, would tackle another hindrance cited by LA county, that is, a lack of sufficient funding. Both these advantage coming in addition to faster and more fitting assignments for all categories of unhoused persons including those left out (like the disabled with project Roomkey) by current assignment procedures.

\par Of course the matching algorithm alone can not fix homelessness and has to be supplemented by already-existing programs for job placement, drug addiction rehabilitation, domestic violence prevention and recovery programs, health care provision, among others. We do not propose this mechanism as a overhauling solution but rather as a more efficient piece to be plugged into the vast effort to end homelessness.

\subsection{Future work}
We hope to obtain the support of many policy-makers and homeless service providers from different cities, support in the form of homelessness data, for example data from the recent project Roomkey effort, and a clear outline of the current housing assignment procedure. This would allow for numerical investigation on the effectiveness of this matching mechanism on real world data. We also intend to go beyond research and actually work with the same city policy makers and homeless service providers in implementing this algorithm in the field. In particular, we are seeking a collaboration with LA county to make this matching mechanism a part of future project Roomkey and Homekey efforts.
\par The unique structure that we presented in section 5 will also be a subject of future work as we look to understand it's mathematical implication and where else we see it in the modern socio-economic societies. One quick example as mentioned earlier is the Food and donations industry.
\section{Acknowledgement}
Dr. Victoria Basolo (UCI), Dr. Amelia Regan (UCI), Samantha A. Carter (UCI), and Sophia G. Bardetti (Middlebury College), thank you all for your wise insight.
\bibliographystyle{bibtex/splncs_srt}
\bibliography{creds}
\end{document}